\documentclass[onecolumn,showkeys,notitlepage]{revtex4-1} 

\usepackage{braket}
\usepackage{graphicx}
\usepackage{amsfonts}
\usepackage{amsmath}
\usepackage{graphicx}
\usepackage{etoolbox}

\begin{document}

\title{Higher-Order Interactions in Quantum Optomechanics: Analysis of Quadratic Terms}

\author{Sina Khorasani} 
\affiliation{Vienna Center for Quantum Science and Technology, University of Vienna, 1090 Vienna, Austria; sina.khorasani@ieee.org}

\begin{abstract}
This article presents a full operator analytical method for studying the quadratic nonlinear interactions in quantum optomechanics. The method is based on the application of higher-order operators, using a six-dimensional basis of second order operators which constitute an exactly closed commutators. We consider both types of standard position-field and the recently predicted non-standard momentum-field quadratic interactions, which is significant when the ratio of mechanical frequency to optical frequency is not negligible. This unexplored regime of large mechanical frequency can be investigated in few platforms including the superconducting electromechanics and simulating quantum cavity electrodynamic circuits. It has been shown that the existence of non-standard quadratic interaction could be observable under appropriate conditions. 
\end{abstract}


\flushbottom
\maketitle
\thispagestyle{empty}

\section*{Introduction}

The field of quantum optomechanics \cite{Kip,Aspel,Bowen} is flourishing as one of the modern applications of quantum physics, where interactions of optical field and mechanical motion inside a confined cavity is being studied. Nonlinear interactions in optomechanics play a critical role in a growing number of studied physical phenomena, which include emergence of second-order mechanical side-bands \cite{q1,q12}, nonlinear optomechanical induced transparency \cite{q2,q3,q4,q8}, phonon-laser \cite{q5,q13}, nonlinear reciprocity \cite{q6}, and parameteric phonon-phonon coupling for cascaded optical transparency \cite{q7}. Recently, existence of optomechanical chaos has been confirmed \cite{q9}, which is contingent on and can be explained only using the nonlinearity of interactions. Meterology with optomechanical symmetry breaking \cite{q10} as well as Kerr-type nonlinear interactions \cite{q8,q11} are other examples of optomechanical phenomena which call for nonlinear analysis. Interestingly, it has been also shown using an extensive analysis, that the gravitational constant can be measured by increased precision using nonlinear optomechanics \cite{q14}, reaching an ideal fundamental sensitivity of $10^{-15}\text{ms}^{-2}$ for state-of-the-art parameters.

Many of the studies in this field utilize linearization of photonic $\hat{a}$ and phononic  $\hat{b}$ ladder operators around their mean values as $\hat{a}\to\bar{a}+\hat{a}$ and $\hat{b}\to\bar{b}+\hat{b}$, where the substituted ladder operators now represent field fluctuations around their respective mean values. This way of linearization is however insufficient for quadratic \cite{Quad1,Quad3,Quad4,Quad5,Quad6,Quad7,Quad8,Bruschi} and higher-order interactions where the resulting Langevin equations \cite{Langevin1,Langevin2,Langevin3,Langevin4} are expected to be strongly nonlinear. Full linearization essentially transforms back every such nonlinear interaction picture into the simple linearized form of $(\hat{a}+\hat{a}^\dagger)(\hat{b}+\hat{b}^\dagger)$. As a result, the behavior of systems under study due to quadratic and higher-order interactions becomes indifferent to that of ordinary linearized optomechanics, implying that in this way of linearization some of the important underlying nonlinear physics may be lost.

Recently, theory of optomechanics has been revisited by the author \cite{Paper1} as well as others for nonlinear \cite{Sala} and quadratic \cite{Thesis} interactions. It has been shown that a non-standard quadratic term could exist due to momentum-field interaction and relativistic effects, the strength of which is proportional to $(\Omega/\omega)^2$ with $\Omega$ and $\omega$ respectively being the mechanical and optical frequencies \cite{Paper1}. Normally, such momentum-field interactions are not expected to survive under the regular operating conditions of large optical frequencies $\omega>>\Omega$. However, it would be a matter of question that whether they could be observable in spectral response of a cavity, given that the $\omega$ and $\Omega$ could be put within the same order of magnitude? The answer is Yes.

Superconducting electromechanics provides a convenient means to observe quadratic effects at such conditions. Not only high photon cavity numbers could be attained, but also mechanical and radio or microwave frequencies could be set within the same order of magnitude with relative ease. This conditions have been actually met at least in one reported experiment \cite{Quad2}, where the mechanical and superconducting circuit frequencies are designed to be equal at $\omega=\Omega=2\pi\times 720\text{kHz}$. Tuning to the lowest-order odd-profiled mechanical mode, or using the membrane-in-the-middle setup \cite{q4,Quad2} in optomechanical experiments could altogether eliminate the standard optomechanical interaction, leaving only the quadratic terms and higher. 

Apart from experimental considerations for observation of quadratic effects, there remains a major obstacle in theoretical analysis of combined standard and non-standard quadratic interactions. To the best knownledge of the author, this regime has been investigated theoretically for the case of only standard quadratic interaction using the time-evolution operators \cite{Bruschi}. 

Growing out of the context of quantum optomechanics, the method of higher-order operators developed by the author \cite{Paper2} can address problems with any general combination of nonlinearity, stochastic input, operator quantities, and spectral estimation. Higher-order operators have been already used in analysis of nonlinear standard optomechanics \cite{Paper3}, where its application has uncovered effects known as sideband inequivalence, quantities such as coherent phonon population, as well as corrections to the optomechanical spring effect, zero-point field optomechanical interactions, and a minimal basis with the highest-order which allows exact integration of optomechanical Hamiltonian subject to multiplicative noise input. Also, it has been independently used \cite{Quad9} for investigation of quadratic effects. But this method has not been verified yet for non-standard quadratic optomechanics, which is the central topic of this study.

In this article, we employ a six-dimensional basis of higher-order operators, all being second order, which satisfy an exact closed commutation relations. This basis can be used to analyze the quadratic interactions of both standard and non-standard types, which has been so far not done. It has been shown that the momentum-field interaction, referred to as the non-standard quadratic term, does have observable effects on the spectral response of the optomechanical cavity, if the design criteria could violate $\Omega<<\omega$. Hence, the effect of this interaction should not be overlooked when the ratio $\Omega/\omega$ is non-negligible. The non-standard term appears to survive even under weak quadratic coupling. This study paves the way for probing a previously unexplored domain of quantum optomechanics.

\section*{Results}
In this section, we discuss the model Hamiltonian for the quadratic interaction in quantum optomechanics. As it will be shown, it is composed of two contributing terms. The first term $\mathbb{H}_{1}$ is the well-known standard quadratic term, resulting from the product of photon number $\hat{n}=\hat{a}^\dagger\hat{a}$ and squared displacement $\mathbb{X}^2=(\hat{b}+\hat{b}^\dagger)^2$. The second term $\mathbb{H}_2$, which is also quadratic in order, represents the non-standard term and describes the momentum-field interaction among the squared momentum of the mirror $\mathbb{P}^2=(\hat{b}-\hat{b}^\dagger)^2$ and squared second quadrature of the electromagnetic field $\mathcal{P}^2=(\hat{a}-\hat{a}^\dagger)^2$ \cite{Paper1}, thus admitting the form $\mathcal{P}^2\mathbb{P}^2$. It should be mentioned that the phase of $\hat{a}$ can be arbitrarily shifted by $\pi/2$, allowing one to rewrite the latter interaction as $\mathcal{X}^2\mathbb{P}^2$, where $\mathcal{X}^2=(\hat{a}+\hat{a}^\dagger)^2$. In that sense, the referral to the field quadratures as either $\mathcal{X}$ or $\mathcal{P}$ is quite arbitrary.

\subsection*{Hamiltonian}

For the purpose of this article, we consider that the standard optomechanical interaction $\mathbb{H}_\text{OM}$ vanishes due to appropriate design with $g_0=0$. This not only simplifies the description of the problem and reduces the dimension of basis significantly, but is also favorable from an experimental point of view, since the lowest order surviving interaction now would be quadratic. As mentioned in the above, this criterion can be easily met in superconducting electromechanics by tuning the electroagnetic frequency to the first odd-profiled mechanical mode, or in optomechanics by using arrangements such as the membrane-in-the-middle setup \cite{q4,Quad2}. In general, the condition $g_0=0$ might not be exactly achieved and it may cause some difficulty in observation of quadratic effects, since standard optomechanical interactions could still mask out the much weaker quadratic interactions. However, for the purpose of current study, we may neglect this effect in the same way is being done by other researchers \cite{Quad1,Quad3,Quad4,Quad5,Quad6,Quad7,Quad8,Bruschi}. The total Hamiltonian is thus given by $\mathbb{H}=\mathbb{H}_\text{0}+\mathbb{H}_1+\mathbb{H}_2+\mathbb{H}_\text{d}$ where
\begin{eqnarray}
\label{eq1}
\mathbb{H}_{0}&=&\hbar\tilde{\omega}\hat{a}^\dagger\hat{a}+\hbar\tilde{\Omega}\hat{b}^\dagger\hat{b}=\hbar\tilde{\omega}\hat{n}+\hbar\tilde{\Omega}\hat{m},\\ \nonumber
\mathbb{H}_{1}&=&\hbar \tfrac{1}{2}\varepsilon\hat{a}^\dagger\hat{a}(\hat{b}+\hat{b}^\dagger)^2=\hbar \tfrac{1}{2}\varepsilon\hat{n}\mathbb{X}^2, \\ \nonumber
\mathbb{H}_{2}&=&-\hbar \tfrac{1}{2}\beta (\hat{a}-\hat{a}^\dagger)^2(\hat{b}-\hat{b}^\dagger)^2=-\hbar \tfrac{1}{2}\beta\mathcal{P}^2\mathbb{P}^2, \\ \nonumber
\mathbb{H}_\text{d}&=&\hbar\gamma(\alpha e^{i\omega_\text{d} t}\hat{a}+\alpha^\ast e^{-i\omega_\text{d} t}\hat{a}^\dagger).
\end{eqnarray}
\noindent
Here, $\tilde{\omega}$ and $\tilde{\Omega}$ are respectively the bare unperturbed optical and mechanical frequencies. We notice that the relative frequency notation of optical detuning, which is useful in standard optomechanical and quadratic interactions \cite{Paper3} is not to be used here. In (\ref{eq1}), furthermore, we assume the existence of only one drive term with the complex amplitude $\alpha$ and frequency $\omega_\text{d}$. In general, it is possible to assume the existence of multiple drive terms at different frequencies, but this does not alter the mathematical approach under consideration. We furthermore notice that the non-standard quadratic term can be also written as $\mathbb{H}_2=-\hbar\tfrac{1}{2}\beta\mathcal{X}^2\mathbb{P}^2$ with basically no physically significant difference, as mentioned in the above section, too. Hence, the non-standard term can take on either of the sign conventions $--$ as $\mathcal{P}^2\mathbb{P}^2$, or $+-$ as $\mathcal{X}^2\mathbb{P}^2$. We shall proceed with the latter. Also, $\varepsilon$ is the strength of the standard quadratic interaction and $\beta$ is the strength of the non-standard quadratic interaction. These are shown to be related as \cite{Paper1}
\begin{equation}
\label{eq2}
\beta=\frac{1}{4}\left(\frac{\pi^2}{3}+\frac{1}{4}\right)\left(\frac{\tilde{\Omega}}{\tilde{\omega}}\right)^2\varepsilon\equiv \tfrac{1}{2}\rho\varepsilon.
\end{equation}
It is easily seen that in the regime of large optical frequency $\tilde{\omega}>>\tilde{\Omega}$, we get $\beta\approx 0$ and the non-standard term vanishes. This is what has actually been probed in nearly all experiments on quadratic optomechanical interactions so far \cite{Quad1,Quad3,Quad4,Quad5,Quad6,Quad7,Quad8,Quad9}. The large mechanical frequency regime of standard quadratic interactions has been however recently probed \cite{Bruschi} and it has been suggested that the roles of optical and mechanical parts are expected to interchange without consideration of the non-standard quadratic effect. Hence, the condition defining the critical mechanical frequency as $\tilde{\Omega}=\frac{1}{2}\sqrt{\frac{1}{3}\pi^2+\frac{1}{4}}\tilde{\omega}\approx 0.941\tilde{\omega}$ marks a critical value for the transition border, across which the regimes of large and small mechanical frequency with respect to the given electromagnetic frequency are distinguished. This happens to be remarkably close to the identical frequencies as $\tilde{\Omega}=\tilde{\omega}$, too. It is not difficult to see that the above Hamiltonian with the sign convention taken as $+-$ can be rewritten as
\begin{equation}
\label{eq3}
\mathbb{H}=\hbar\omega\hat{n}+\hbar{\Omega}\hat{m}+\hbar\epsilon\left[\hat{n}(\hat{m}+\hat{d}+\hat{d}^\dagger)-\rho(\hat{n}+\hat{c}+\hat{c}^\dagger)(\hat{d}+\hat{d}^\dagger-\hat{m})\right]+\mathbb{H}_\text{d},
\end{equation}
\noindent
in which $\hat{c}=\frac{1}{2}\hat{a}^2$ and $\hat{d}=\frac{1}{2}\hat{b}^2$ are defined and discussed extensively in the preceding articles \cite{Paper1,Paper2,Paper3}, and time-independent non-interacting terms are dropped which are irrelevant to the behavior of system dynamics. Furthermore, the altered effective optical and mechanical frequencies due to the quadratic optomechanical interaction assume different forms, and now read
\begin{equation}
\label{eq4}
\omega=\tilde{\omega}+\tfrac{1}{2}\varepsilon+\beta, \quad \Omega=\tilde{\Omega}+\beta.
\end{equation}

The importance of the non-standard quadratic term $\mathbb{H}_2$ \cite{Paper1} is that it describes a non-vanishing correction to the field-mirror interaction, beyond simple nonlinear quantum back-action of mirror on the reflected light. In standard picture of quantum optomechanics, the light gets reflected off a displaced mirror which already has shifted the resonance frequency of cavity. Higher-order corrections to combination of these effects give rise to quadratic interactions. But in quadratic quantum optomechanics, either momenta of field and mirror do not apparently come into consideration, or their contributions are somehow lost because of the approximations used in the expansions. It is expected that normally such an interaction should take care of momentum exchange. The exchange and conservation of momenta under standard quadratic interaction is uncertain and actually not quite obvious, since momentum operators do not show up in the interaction. 

\subsection*{Basis}

Analysis of the Hamiltonian (\ref{eq3}) here is going to be based on the six-dimensional space spanned by the basis operators $\{A\}^\text{T}=\{\hat{c},\hat{c}^\dagger,\hat{n},\hat{d},\hat{d}^\dagger,\hat{m}\}$ \cite{Paper2}. This basis can be easily seen to be the smallest possible set, with closed commutators, capable of describing the system modeled by (\ref{eq1}). The commutation properties of this basis \cite{Paper2} is here given for the sake of convenience
\begin{eqnarray}
\label{eq5}
[\hat{c},\hat{c}^\dagger]=\hat{n}+\tfrac{1}{2},&\quad
[\hat{c},\hat{n}]=2\hat{c}, &\quad
[\hat{n},\hat{c}^\dagger]=2\hat{c}^\dagger, \\ \nonumber
[\hat{d},\hat{d}^\dagger]=\hat{m}+\tfrac{1}{2},&\quad
[\hat{d},\hat{m}]=2\hat{d}, &\quad
[\hat{m},\hat{d}^\dagger]=2\hat{d}^\dagger.
\end{eqnarray}
\noindent
All commutators among photonic $\{\hat{c},\hat{c}^\dagger,\hat{n}\}$ and phononic operators $\{\hat{d},\hat{d}^\dagger,\hat{m}\}$ are clearly zero. This fact together with the set of commutators (\ref{eq5}) establishes the closedness property of our basis. The basis $\{A\}$ under consideration is called to be second-order, since its operators are all products of two single ladder operators. Other bases of the third- and higher-orders are discussed elsewhere, respectively for standard optomechanics \cite{Paper3} and quadratic interactions \cite{Paper2}.

Furthermore, we will need $[\hat{c},\hat{a}]=[\hat{c}^\dagger,\hat{a}^\dagger]=0$ along with the pair of relationships
\begin{equation}
\label{eq6}
[\hat{c},\hat{a}^\dagger]=[\hat{a},\hat{n}]=\hat{a},\quad 
[\hat{a},\hat{c}^\dagger]=[n,\hat{a}^\dagger]=\hat{a}^\dagger, 
\end{equation}
to evaluate the effect of drive and input noise terms later. Assumption of a resonant drive term here requires $\omega_\text{d}=\omega$, which highlights a constant shift in cavity optical frequency because of the presence of quadratic optomechanical interactions. This will greatly simplify the mathematics involved in construction of Langevin equations. These latter relations (\ref{eq5},\ref{eq6}) show that the eight-dimensional basis $\{\hat{a},\hat{a}^\dagger\}\bigcup\{A\}$, which is of the mixed first and second-order, also constitutes a closed commutation relationships. However, for the purpose of this article, the original six-dimensional basis $\{A\}$ is sufficient.

\subsection*{Langevin Equations}

The task of construction of Langevin equations proceeds with the original equation \cite{Kip,Aspel,Bowen,Langevin1,Langevin2,Langevin3,Langevin4}
\begin{equation}
\label{eq7}
\dot{\hat{z}}=-\frac{i}{\hbar}[\hat{z},\mathbb{H}]-[\hat{z},\hat{x}^\dagger]\left(\tfrac{1}{2}\gamma\hat{x}+\sqrt{\gamma}\hat{x}_\text{in}\right)+\left(\tfrac{1}{2}\gamma\hat{x}^\dagger+\sqrt{\gamma}\hat{x}_\text{in}^\dagger\right)[\hat{z},\hat{x}],
\end{equation}
\noindent
where $\hat{z}$ is an arbitrary operator belonging to the basis, $\hat{x}$ is any system annihilator associated with the decay rate $\gamma$, and $\hat{x}_\text{in}$ is the corresponding input field including contributions from both of the deterministic drive and stochastic noise terms. Also, we may set either $\hat{x}=\hat{a}$ with $\gamma=\kappa$ and $\hat{x}_\text{in}=\hat{a}_\text{in}$, or $\hat{x}=\hat{b}$ with $\gamma=\Gamma$ and $\hat{x}_\text{in}=\hat{b}_\text{in}$ to allow simple and straightforward construction of noise terms. This particular choice avoids appearance of squared noises, which are otherwise required \cite{Paper2}.

There are a total of six Langevin equations for the system (\ref{eq1}), which can be constructed one by one, corresponding to the six members of $\{A\}$. These equations can be obtained using the commutators (\ref{eq5},\ref{eq6}) after some straightforward algebra as
\begin{eqnarray}
\label{eq8}
\dot{\hat{c}}&=&-i2\omega\hat{c}-\kappa \hat{c}-i\varepsilon\left[2(\hat{m}+\hat{d}+\hat{d}^\dagger)\hat{c}-\rho(\hat{d}+\hat{d}^\dagger-\hat{m})(2\hat{c}+\hat{n}+\tfrac{1}{2})\right]+i\alpha^\ast e^{-i\omega t}\hat{a}-\sqrt{\kappa}\hat{a}\hat{a}_\text{in} ,\\ \nonumber
\dot{\hat{c}}^\dagger&=&i2\omega\hat{c}^\dagger-\kappa \hat{c}^\dagger+i\varepsilon\left[2(\hat{m}+\hat{d}+\hat{d}^\dagger)\hat{c}^\dagger-\rho(\hat{d}+\hat{d}^\dagger-\hat{m})(2\hat{c}^\dagger+\hat{n}+\tfrac{1}{2})\right]-i\alpha e^{i\omega t}\hat{a}^\dagger-\sqrt{\kappa}\hat{a}_\text{in}^\dagger\hat{a}^\dagger ,\\ \nonumber
\dot{\hat{n}}&=&-i2\beta(\hat{c}-\hat{c}^\dagger)(\hat{d}+\hat{d}^\dagger-\hat{m})-\kappa\hat{n}+i(\alpha e^{i\omega t}\hat{a}-\alpha^\ast e^{-i\omega t}\hat{a}^\dagger)-\sqrt{\kappa}(\hat{a}^\dagger\hat{a}_\text{in}+\hat{a}_\text{in}^\dagger\hat{a}), \\ \nonumber
\dot{\hat{d}}&=&-i2\Omega\hat{d}-\Gamma \hat{d}-i\varepsilon\left[\hat{n}(2\hat{d}+\hat{m}+\tfrac{1}{2})-\rho(\hat{c}+\hat{c}^\dagger+\hat{n})(\hat{m}+\tfrac{1}{2}-2\hat{d})\right]-\sqrt{\Gamma}\hat{b}\hat{b}_\text{in} ,\\ \nonumber
\dot{\hat{d}}^\dagger&=&i2\Omega\hat{d}^\dagger-\Gamma \hat{d}^\dagger+i\varepsilon\left[\hat{n}(2\hat{d}^\dagger+\hat{m}+\tfrac{1}{2})-\rho(\hat{c}+\hat{c}^\dagger+\hat{n})(\hat{m}+\tfrac{1}{2}-2\hat{d}^\dagger)\right]-\sqrt{\Gamma}\hat{b}_\text{in}^\dagger\hat{b}^\dagger ,\\ \nonumber
\dot{\hat{m}}&=&-i2\varepsilon(\hat{d}-\hat{d}^\dagger)\left[(\rho-1)\hat{n}+\rho(\hat{c}+\hat{c}^\dagger)\right]-\sqrt{\Gamma}(\hat{b}^\dagger\hat{b}_\text{in}+\hat{b}_\text{in}^\dagger\hat{b}).
\end{eqnarray}

\subsection*{Steady-State Equilibrium}
The special algebraic form of non-standard interaction together with the presence of drive term makes the analysis requiring a bit of care. Ambiguities could be avoided by using explicit decompositions and replacements
\begin{eqnarray}
\label{eq9}
\hat{a}(t)\to\bar{a}e^{-i\omega t}+\hat{a}(t), &\quad&
\hat{a}^\dagger(t)\to\bar{a}^\ast e^{i\omega t}+\hat{a}^\dagger(t), \\ \nonumber
\hat{n}(t)\to\bar{n}+\hat{n}(t), &\quad&
\hat{m}(t)\to\bar{m}+\hat{m}(t),
\end{eqnarray}
\noindent
where $\bar{n}=|\bar{a}|^2$ is the average cavity photon number and $\bar{m}$ is the average cavity phonon number. From now on, $\hat{n}$ and $\hat{m}$ will represent only deviations from the average steady-state populations. In a similar manner, we can employ the decompositions and replacements
\begin{eqnarray}
\label{eq10}
\hat{c}(t)\to\bar{c}e^{-2i\omega t}+\hat{c}(t), &\quad&
\hat{c}^\dagger(t)\to\bar{c}^\ast e^{2i\omega t}+\hat{c}^\dagger(t), \\ \nonumber
\hat{d}(t)\to\bar{d}e^{-2i\Omega t}+\hat{d}(t), &\quad&
\hat{d}^\dagger(t)\to\bar{d}^\ast e^{2i\Omega t}+\hat{d}^\dagger(t).
\end{eqnarray}
\noindent
It has to be mentioned that the replacements (\ref{eq9},\ref{eq10}) are not linearization, but rather an algebraic convention which further allows distinction of steady-state cavity photon and phonon populations under resonant drive. The first separated scalar terms represent the major non-oscillating parts $\bar{n}$ and $\bar{m}$, and oscillating parts for the rest.

One may now proceed to construct the equations for steady-state populations $\bar{n}$ and $\bar{m}$. This requires employing the replacements (\ref{eq9},\ref{eq10}) in Langevin equations (\ref{eq8}), taking the derivatives on the left, multiplying both sides of the first and fourth equations of (\ref{eq8}) by respectively $e^{i2\omega t}$ and $e^{i2\Omega t}$, and discarding all remaining time-dependent terms. This is equivalent to a Rotating-Wave Approximation (RWA) analysis, but carried out at double frequencies. Obviously, all subsequent derivations and calculations in the steady-state will retain their validity within the constraint of RWA condition.

Under resonance conditions where optical and mechanical frequencies are the same $\omega=\pm\Omega$, there remain a few extra terms. After discarding stochastic noise input, application of these steps to the first, third, fourth, and sixth equations of (\ref{eq8}) and simplifying, gives the following four nonlinear algebraic equations for steady state terms
\begin{eqnarray}
\label{eq11}
i\alpha^\ast\sqrt{\bar{n}}&=&\tfrac{1}{2}\kappa\bar{n}+i\varepsilon\left[\tfrac{1}{2}(1+\rho)\bar{m}\bar{n}-\rho(\bar{n}+\tfrac{1}{2})(\delta_{\omega,\Omega}\bar{d}+\delta_{\omega,-\Omega}\bar{d}^\ast)\right],\\ \nonumber
\bar{m}^2-\bar{m}&=&4|\bar{d}|^2,\\ \nonumber
-\Gamma\bar{d}&=&i\varepsilon\left[2(1+\rho)\bar{n}\bar{d}-\rho(\bar{m}+\tfrac{1}{2})(\delta_{\omega,\Omega}+\delta_{\omega,-\Omega})\bar{n}\right], \\ \nonumber
\kappa\bar{n}&=&i(\alpha-\alpha^\ast)\sqrt{\bar{n}}-i2\beta(\bar{d}^\ast-\bar{d})\bar{n}(\delta_{\omega,\Omega}-\delta_{\omega,-\Omega}).
\end{eqnarray}
\noindent
Here, $\delta_{\theta,\varphi}$ represents the Kronecker's delta, and equals $1$ if $\theta=\varphi$ and $0$ otherwise. Also, $\alpha$ is the complex drive amplitude as already defined in the above under (\ref{eq1}). These are four equations in terms of $\bar{n}$, $\bar{m}$, $\bar{d}$, and $\angle\alpha$, and we notice that $|\alpha|=\eta P_\text{op}/\hbar\omega$ is the photon flux incident on the cavity due to the optical power $P_\text{op}$, where $\eta$ is the input coupling efficiency. When there is no resonance between optics and mechanics with $\delta_{\pm\omega,\pm\Omega}=0$, the system (\ref{eq11}) reduces to the fairly simple pair 
\begin{equation}
\label{eq12}
|\alpha|^2=\varepsilon^2(1+\rho)^2\bar{m}^2\bar{n}, \quad
\bar{n}|\alpha|^2=\varepsilon^2(\bar{m}^2-\bar{m}).
\end{equation}
These two nonlinear algebraic equations can be numerically solved for non-negative real roots of $\bar{n}$ and $\bar{m}$.

\subsection*{Integrable System \& Stability}

The Langevin equations (\ref{eq8}) are still nonlinear and thus non-integrable, but they are instead expressed in terms of second-order operators. Hence, even after taking out the constant oscillating parts using the replacements (\ref{eq9},\ref{eq10}), and ignoring the remaining fourth- and higher-order nonlinear terms, the basic nonlinear quadratic interaction still survives. This advantage in using higher-order operators has been noticed by the author \cite{Paper2} as well as others \cite{Quad9}. 

Following the one-dimensional analysis provided in the preceding study \cite{Paper1}, the strength of subsequent nonlinear nth-order terms decrease typically as $(x_\text{zp}/l)^n$ where $x_\text{zp}$ is the mechanical zero-point fluctuations of vacuum and $l$ is the typical length scale of the cavity. Having that said, even and odd powers grow almost proportionally. So, once the membrane-in-the-middle or any equivalent setup which allows vanishing $g_0$ is employed, there would be no non-zero odd-order optomechanical interaction at all. Similarly, the lowest non-vanishing nonlinear term after the quadratic term would be the sextic (6th) order interaction which is expected to be weaker by a factor of $(x_\text{zp}/l)^n$ at least. This ratio shall be normally too small to be of any physical significance. Doing this leaves the linearized integrable form
\begin{eqnarray}
\label{eq13}
\dot{\hat{c}}&=&[i2\omega-\kappa-i2(\varepsilon+\beta)\bar{m}]\hat{c}+i\beta(\bar{n}+\tfrac{1}{2})(\hat{d}+\hat{d}^\dagger-\hat{m})-i\beta\bar{m}\hat{n}-\sqrt{\tfrac{1}{2}\bar{n}\kappa}\hat{a}_\text{in}, \\ \nonumber
\dot{\hat{c}}^\dagger&=&[-i2\omega-\kappa+i2(\varepsilon+\beta)\bar{m}]\hat{c}^\dagger-i\beta(\bar{n}+\tfrac{1}{2})(\hat{d}+\hat{d}^\dagger-\hat{m})+i\beta\bar{m}\hat{n}-\sqrt{\tfrac{1}{2}\bar{n}\kappa}\hat{a}_\text{in}^\dagger, \\ \nonumber
\dot{\hat{n}}&=&-\kappa\hat{n}+i2\beta\bar{m}(\hat{c}-\hat{c}^\dagger)-\sqrt{\kappa\bar{n}}(\hat{a}_\text{in}+\hat{a}_\text{in}^\dagger),\\ \nonumber
\dot{\hat{d}}&=&[-i2\Omega-\Gamma-i2(\varepsilon+\beta)\bar{n}]\hat{d}+i\beta(\bar{m}+\tfrac{1}{2})(\hat{c}+\hat{c}^\dagger)-i(\varepsilon-\beta)[\bar{n}\hat{m}+(\bar{m}+\tfrac{1}{2})\hat{n}]-\sqrt{2\Gamma|\bar{d}|}\hat{b}_\text{in},\\ \nonumber
\dot{\hat{d}}^\dagger&=&[i2\Omega-\Gamma+i2(\varepsilon+\beta)\bar{n}]\hat{d}^\dagger-i\beta(\bar{m}+\tfrac{1}{2})(\hat{c}+\hat{c}^\dagger)+i(\varepsilon-\beta)[\bar{n}\hat{m}+(\bar{m}+\tfrac{1}{2})\hat{n}]-\sqrt{2\Gamma|\bar{d}|}\hat{b}_\text{in}^\dagger,\\ \nonumber
\dot{\hat{m}}&=&-\Gamma\hat{m}+i2(\varepsilon-\beta)\bar{n}(\hat{d}-\hat{d}^\dagger)-\sqrt{\Gamma|\bar{d}|}(\hat{b}_\text{in}+\hat{b}_\text{in}^\dagger).
\end{eqnarray} 
\noindent
Here, we have furthermore employed the white noise approximation to the Weiner processes $\hat{a}_\text{in}$ and $\hat{b}_\text{in}$. This Markovian approximation makes these noise processes insensitive to any frequency shift, or multiplication by purely oscillating terms such as $\exp(\pm i\omega t)$ and $\exp(\pm i\Omega t)$. This fact facilitates the construction of noise processes, avoiding the burden of higher-order noise terms. Otherwise, terms such as $\hat{a}_\text{in}^2$ and $\hat{b}_\text{in}^2$ enter the Langevin equations \cite{Paper2} which need a careful and very special treatment to evaluate their corresponding spectral densities. Now, the system of Langevin equations (\ref{eq13}) is fully linearized and can be integrated to obtain the spectral densities of each variable. In order to do this, we first define the input vector
\begin{equation}
\label{eq14}
\{A_\text{in}\}^\text{T}=\left\{
\hat{a}_\text{in},
\hat{a}_\text{in}^\dagger, 
\tfrac{1}{2}(\hat{a}_\text{in}+\hat{a}_\text{in}^\dagger), 
\hat{b}_\text{in},
\hat{b}_\text{in}^\dagger,
\tfrac{1}{2}(\hat{b}_\text{in}+\hat{b}_\text{in}^\dagger) 
\right\},
\end{equation}
\noindent
and the diagonal matrix
\begin{equation}
\label{eq15}
[\gamma]=\text{Diag}\left[\bar{n}\kappa,\bar{n}\kappa,4\bar{n}\kappa,2|\bar{d}|\Gamma,2|\bar{d}|\Gamma,4|\bar{d}|\Gamma\right].
\end{equation}
\noindent
The Langevin equations now read
\begin{equation}
\label{eq16}
\frac{d}{dt}\{A\}=[\textbf{M}]\{A\}-\sqrt{[\gamma]}\{A_\text{in}\},
\end{equation}
\noindent
where the coefficients matrix $[\textbf{M}]$ is given as
\begin{footnotesize}
	\begin{equation}
	\label{eq17}
	[\textbf{M}]= 
	\left[
	\begin{array}{cccccc}
	i2(\omega-\zeta\bar{m})-\kappa & 0 & -i\beta\bar{m} & i\beta(\bar{n}+\frac{1}{2}) & i\beta(\bar{n}+\frac{1}{2}) & -i\beta(\bar{n}+\frac{1}{2}) \\ 
	0 & -i2(\omega-\zeta\bar{m})-\kappa & i\beta\bar{m} & -i\beta(\bar{n}+\frac{1}{2}) & -i\beta(\bar{n}+\frac{1}{2}) & i\beta(\bar{n}+\frac{1}{2}) \\ 
	i2\beta\bar{m} & -i2\beta\bar{m} & -\kappa & 0 & 0 & 0 \\
	i\beta(\bar{m}+\frac{1}{2}) & i\beta(\bar{m}+\frac{1}{2}) & -i\chi(\bar{m}+\frac{1}{2}) & -i2(\Omega+\zeta\bar{n})-\Gamma & 0 & -i\chi\bar{n} \\
	-i\beta(\bar{m}+\frac{1}{2}) & -i\beta(\bar{m}+\frac{1}{2}) & i\chi(\bar{m}+\frac{1}{2}) & 0 & i2(\Omega+\zeta\bar{n})-\Gamma & i\chi\bar{n} \\
	0 & 0 & 0 & i2\chi\bar{n} & -i2\chi\bar{n} & -\Gamma
	\end{array}
	\right],
	\end{equation}
\end{footnotesize}
\noindent
Here, $\zeta=\varepsilon+\beta$ and $\chi=\varepsilon-\beta$. This system is fully integrable and linearized, describing a nonlinear Hamiltonian, and can be Fourier transformed to find the spectral densities. Dynamical stability can be easily determined by inspection of the loci of eigenvalues of (\ref{eq17}). The optomechanical system is unconditionally stable if all six eigenvalues have negative real values. 

It is remarkable that any linearization on the operators $\hat{a}$ and $\hat{b}$ done on the Hamiltonian (\ref{eq1}) shall put the interaction in the well recognized forms of either $\mathcal{X}\mathbb{X}$ or $\mathcal{P}\mathbb{P}$, which reverts back to the basic conventional physics of optomechanics \cite{Kip,Aspel,Bowen}. Hence, any linearization process on the first-order operators will eventually discard the physics of higher-order interactions. The method of higher-order operators discussed in this article and previous researches \cite{Paper2,Paper3,Quad9} is targeted to alleviate this problem.

\subsection*{Spectral Density}

The input-output relations \cite{Langevin1,Langevin2,Langevin3} connects the incident and reflection waves of a weak probe beam at frequency $w$ (not to be confused with optical frequency $\omega$). This relationship in Fourier domain together with (\ref{eq16}) yield
\begin{eqnarray}
\label{eq18}
\{A_\text{out}(w)\}&=&\{A_\text{in}(w)\}+\sqrt{[\gamma]}\{A(w)\},\\ \nonumber
\{A_\text{out}(w)\}&=&[\textbf{S}(w)]\{A_\text{in}(w)\},\\ \nonumber
[\textbf{S}(w)]&=&[\textbf{I}]-\sqrt{[\gamma]}\left([\textbf{M}]-iw [\textbf{I}]\right)\sqrt{[\gamma]},
\end{eqnarray}
\noindent
where $[\textbf{I}]$ is the identity matrix. This can be now summarized to yield the spectral density of $\hat{c}$ for positive frequencies $w>0$ as
\begin{equation}
\label{eq19}
S_{CC}(w)=|S_{11}(w)|^2+|S_{13}(w)|^2\tfrac{1}{2}+|S_{14}(w)|^2(\bar{m}+1)+|S_{15}(w)|^2\bar{m}+|S_{14}(w)|^2(\bar{m}+\tfrac{1}{2}),
\end{equation}
\noindent
where $S_{ij}(w)$ are frequency dependent elements of the $6\times 6$ scattering matrix $[\textbf{S}(w)]$ defined in (\ref{eq18}). In the above, we have made use of the fact that spectral density of the input vector $\{\hat{A}_\text{in}(w)\}$ (\ref{eq14}) is 
\begin{eqnarray}
\label{eq20}
\{ S_\text{in}(w>0)\}^{\rm T}&=&\{1,0,\tfrac{1}{2},\bar{m}+1,\bar{m},\bar{m}+\tfrac{1}{2}\}, \\ \nonumber
\{ S_\text{in}(w<0)\}^{\rm T}&=&\{0,1,\tfrac{1}{2},\bar{m},\bar{m}+1,\bar{m}+\tfrac{1}{2}\}.
\end{eqnarray}
\noindent
There are quite a few assumptions needed to obtain (\ref{eq19}), including validity of Gaussian white noise processes for both photons and phonons, complete independence of stochastic processes for phonons and photons, and ignoring higher-order noise terms arising from multiplicative noise terms in (\ref{eq8}) leading to the mean-field approximation for nonlinear multiplicative noise. 

All remains now is to recover the spectral density $S_{AA}(w)$ of the first-order operator $\hat{a}$ from the calculated spectral density $S_{CC}(w)$ of the second-order operator. Assuming that $S_{CC}(w)$ is composed of well isolated Gaussian peaks as $S_{CC}(w)=\sum s_j \exp[-(w-\omega_j)^2/\Delta\omega_j^2]$ with $s_j$ and $\Delta\omega_j$ respectively being the amplitude and spread of the $j$-th Lorentzian, $S_{AA}(w)$ can be related \cite{Paper2} to $S_{CC}(w)$ roughly as
\begin{equation}
\label{eq21}
S_{AA}(w)\approx 2\pi\sqrt{\pi}\sum_j \frac{s_j\omega_j^2}{\Delta\omega_j} \exp\left[-\frac{\left(w-\frac{1}{2}\omega_j\right)^2}{2\Delta\omega_j^2}\right].
\end{equation}
\noindent
Therefore, the spectral density of the requested system operator can be approximately recovered from the calculated spectral density of the higher-order system operator. A factor 4 has been already been absorbed in (\ref{eq21}) because of the definition $\hat{c}=\frac{1}{2}\hat{a}^2$.

\section*{Discussion}

In this section, we discuss numerical results of solving the system (\ref{eq16}) using the scattering matrix formalism (\ref{eq18}) and present the computed spectral densities. We base the simulation on the parameters mostly taken from a recent study on superconducting electromechanics \cite{Quad2} where radio-and mechanical frequencies are accurately tuned and set to equal values. However, we study resonant $\omega=\Omega$, near-resonant $\omega\approx\Omega$, and off-resonant cases $\omega>\Omega$ and $\omega<\Omega$, in weak, strong, and ultrastrong coupling regimes. The cases of strong and ultrastrong coupling are here somehow idealized, and there might be difficulty in attaining those conditions under practical experimental situations. Nevertheless, these can be obtained using appropriately designed circuit cavity-electrodynamics quantum simulation of quadratic optomechanics \cite{QuadCir}.

Since no quadratic interaction is investigated in the article under consideration, we proceed to assume that the interaction with the second mechanical mode is being studied so that $g_0$ may be effectively set to zero. Furthermore, $\varepsilon$ is taken to be one to two orders of magnitude below the typical measured $g_0$ in the system, where $g_0$ is the single-photon optomechanical interaction rate. Simulation parameters are given in Table \ref{Table1}. The numbers in Table \ref{Table1} are already chosen as similar as possible to the experiment \cite{Quad2} where available. This is the only known experimental configuration to the author, where mechanical frequency $\Omega$ and electromagnetic circuit frequency $\omega$ are designed to be within the same order of magnitude, here identical at $\omega=\Omega=2\pi\times 720\text{kHz}$, using a carefully designed membrane-in-the-middle setup. Having that said, not only it is actually possible to probe the regime of large mechanical frequency and even $\omega=\Omega$, as already has been shown, but also a membrane-in-the-middle configuration could provide direct access to quadratic interactions.

\begin{table}[]
	\centering
	\caption{Numerical values of resonant studied superconducting electromechanical system parameters with quadratic interaction, in weak coupling regime. Numbers are taken where available from an experiment \cite{Quad2} with membrane-in-the-middle design. In non-resonant cases, $\omega$ is varied while $\Omega$ is kept constant. For strong and ultrastrong couplings, $\varepsilon$ is increased respectively by a factor of $10^2$ and $10^3$.}
	\label{Table1}
	\begin{tabular}{ccccccc}
		\hline\hline
		$\omega$ & $\Omega$ & $\kappa$ & $\Gamma$ & $T$ & $\varepsilon$  \\ \hline 
		$2\pi\times 720{\rm kHz}$ & $2\pi\times 720{\rm kHz}$ & $2\pi\times 5.5{\rm kHz}$ & $2\pi\times 2.4{\rm Hz}$ & $40{\rm mK}$ & $2\pi\times 5{\rm Hz}$   \\ \hline\hline
	\end{tabular}
\end{table}

\begin{figure}[ht!]
	\centering
	\includegraphics[width=4.8in]{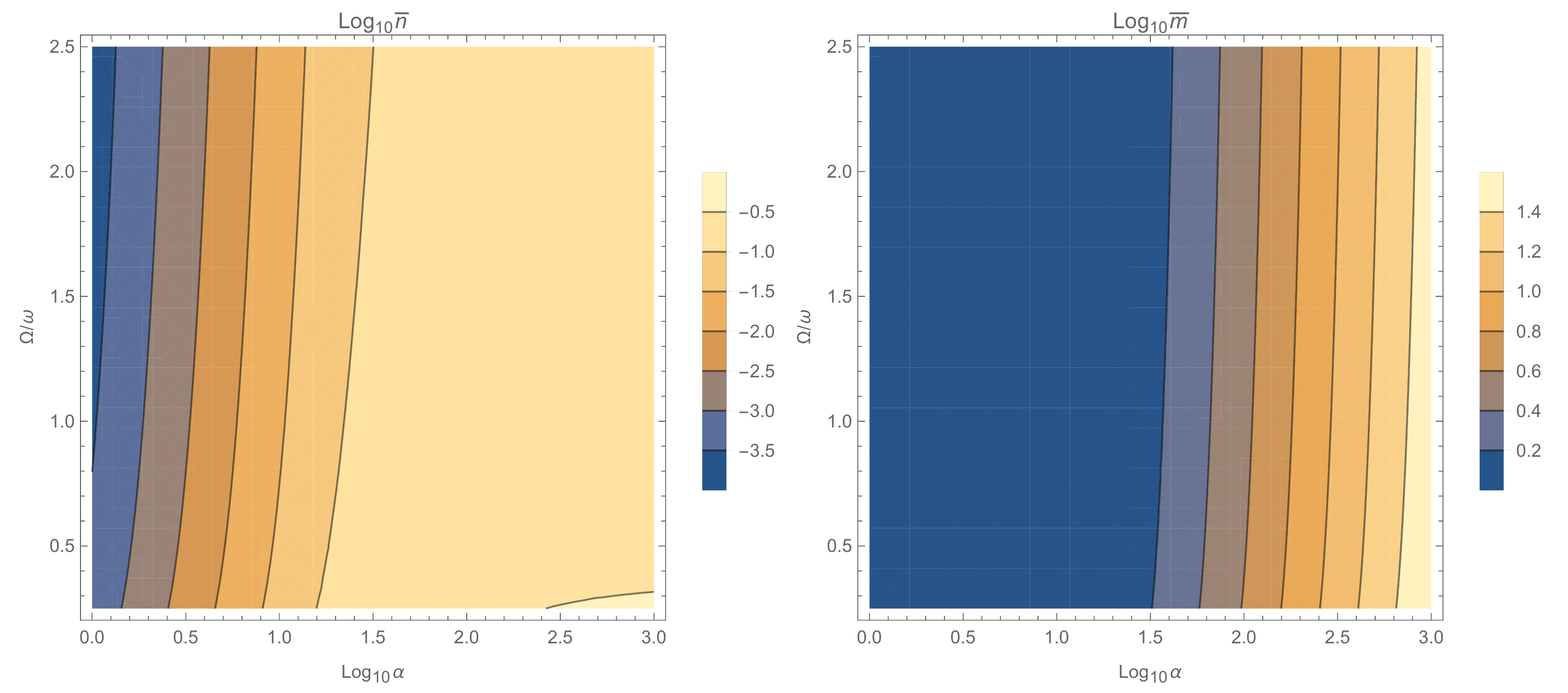}
	\caption{The steady-state photon $\bar{n}$ and phonon $\bar{m}$ population in logarithmic scale, versus input photon flux $\alpha$ and for various mechanical to optical frequency ratios $\Omega/\omega$.\label{Fig1ab}}
\end{figure}

\begin{figure}[ht!]
	\centering
	\includegraphics[width=6.6in]{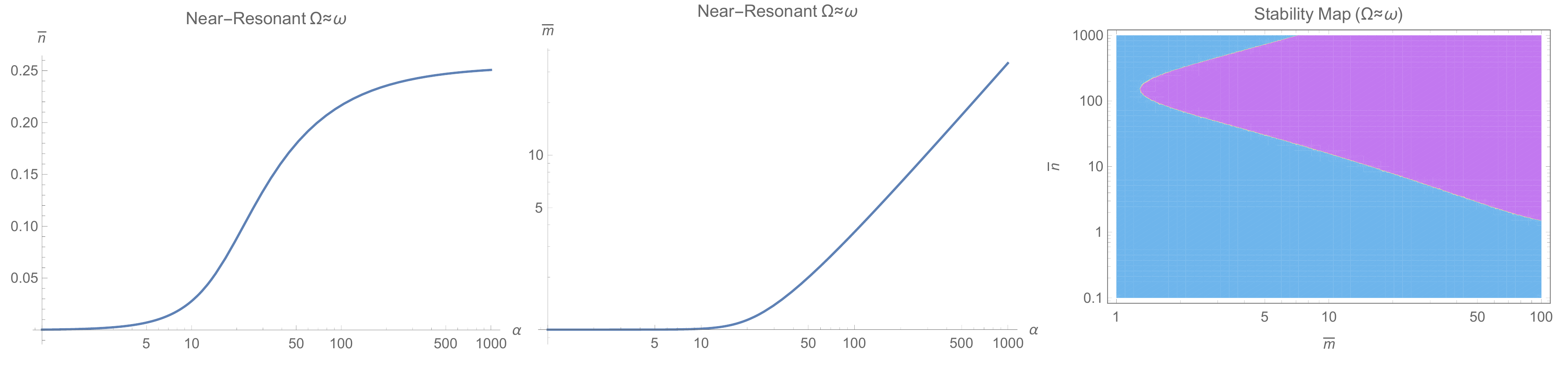}
	\caption{The steady-state photon $\bar{n}$ and phonon $\bar{m}$ population in logarithmic scale, versus input photon flux $\alpha$ at the near-resonant case $\Omega\approx\omega$. The right figure shows the stability diagram in terms of $(\bar{n},\bar{m})$ where cyan color represents dynamically stable domains. For the range $(\bar{n},\bar{m})$ on the left, system is unconditionally stable. \label{Fig1cd}}
\end{figure}

\begin{figure}[ht!]
	\centering
	\includegraphics[width=6.6in]{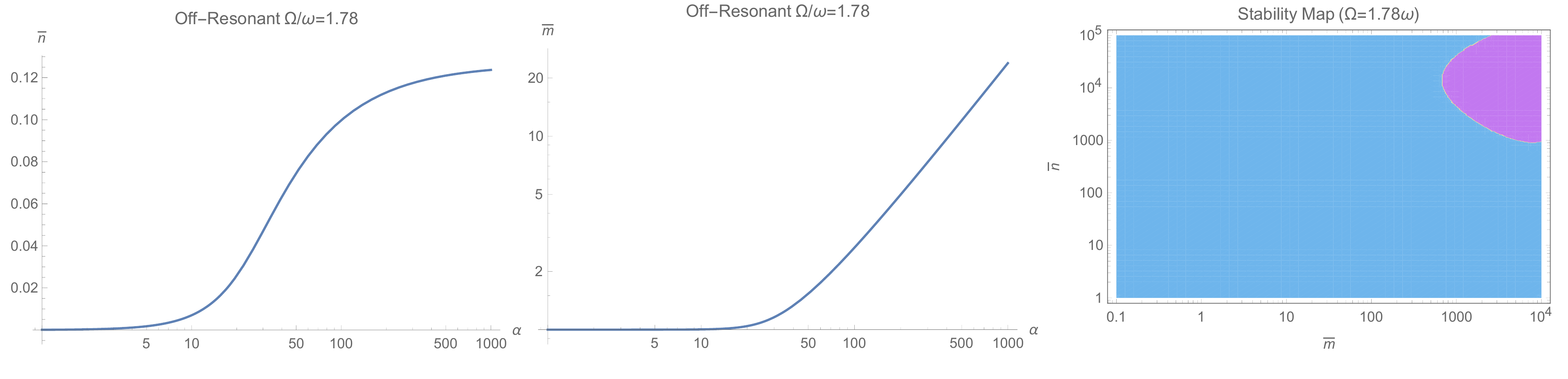}
	\caption{The steady-state photon $\bar{n}$ and phonon $\bar{m}$ population in logarithmic scale, versus input photon flux $\alpha$ at the off-resonant case $\Omega=1.78\omega$. The right figure shows the stability diagram in terms of $(\bar{n},\bar{m})$ where cyan color represents dynamically stable domains. For the range $(\bar{n},\bar{m})$ on the left, system is unconditionally stable. \label{Fig1ef}}
\end{figure}

\begin{figure}[ht!]
	\centering
	\includegraphics[width=6.6in]{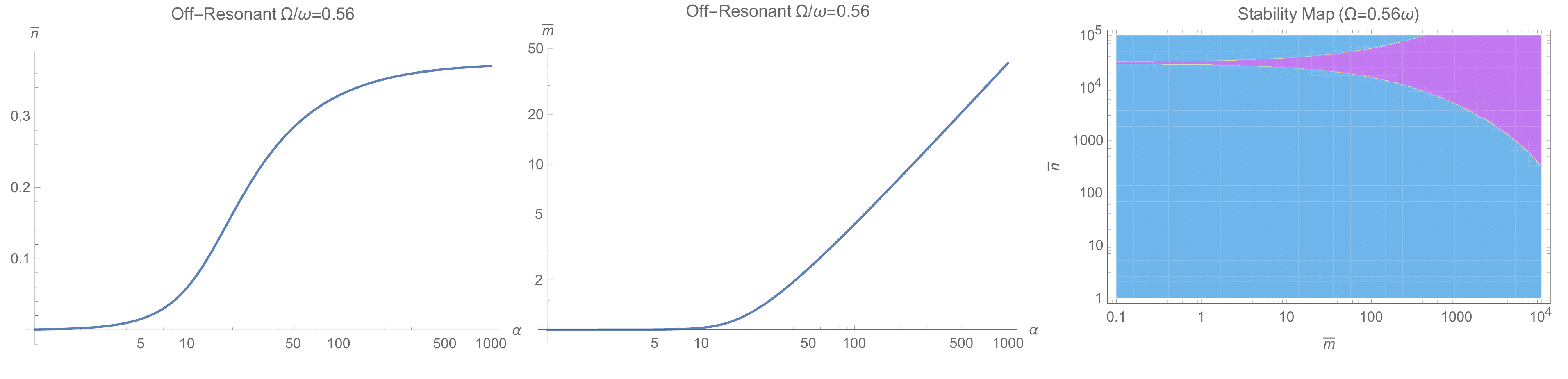}
	\caption{The steady-state photon $\bar{n}$ and phonon $\bar{m}$ population in logarithmic scale, versus input photon flux $\alpha$ at the off-resonant case $\Omega=0.56\omega$. The right figure shows the stability diagram in terms of $(\bar{n},\bar{m})$ where cyan color represents dynamically stable domains. For the range $(\bar{n},\bar{m})$ on the left, system is unconditionally stable. \label{Fig1gh}}
\end{figure}

The steady-state populations of photon $\bar{n}$ and phonons $\bar{m}$ are the first step to solve for. This can be done by numerical solution of (\ref{eq12}) for the near-resonant and non-resonant cases $\omega\neq\Omega$, while for the resonant case with $\omega=\Omega$, the set of equations (\ref{eq11}) has to be solved for positive real roots. The general behavior is that both $\bar{n}$ and $\bar{m}$ increase with input photon flux $\alpha$ as illustrated in Fig. \ref{Fig1ab} as logarithmic contour plots for various frequency ratios $\Omega/\omega$ and input fluxes $\alpha$. However, making ordinary logarithmic plots at near-resonance $\omega\approx\Omega$ in Fig. \ref{Fig1cd}, and off-resonant cases with $\Omega=1.78\omega$ and $\Omega=0.56\omega$ respectively in Figs. \ref{Fig1ef} and \ref{Fig1gh} reveals that both populations $\bar{n}$ and $\bar{m}$  start to increase beyond a certain threshold. In all cases, under weak coupling with $\varepsilon=2\pi\times 5{\rm Hz}$, this threshold is around $\alpha_\text{sat}=20\text{Hz}$. While $\bar{n}$ effectively saturates at high illumination intensities for all non-resonant cases $\omega\neq\Omega$, $\bar{m}$ tends to increase indefinitely. This behavior due to quadratic interaction is remarkably different from what one would expect in standard optomechanics, where $\bar{m}\propto\bar{n}^2$ \cite{Paper3}. 

\begin{figure}[ht!]
	\centering
	\includegraphics[width=6.6in]{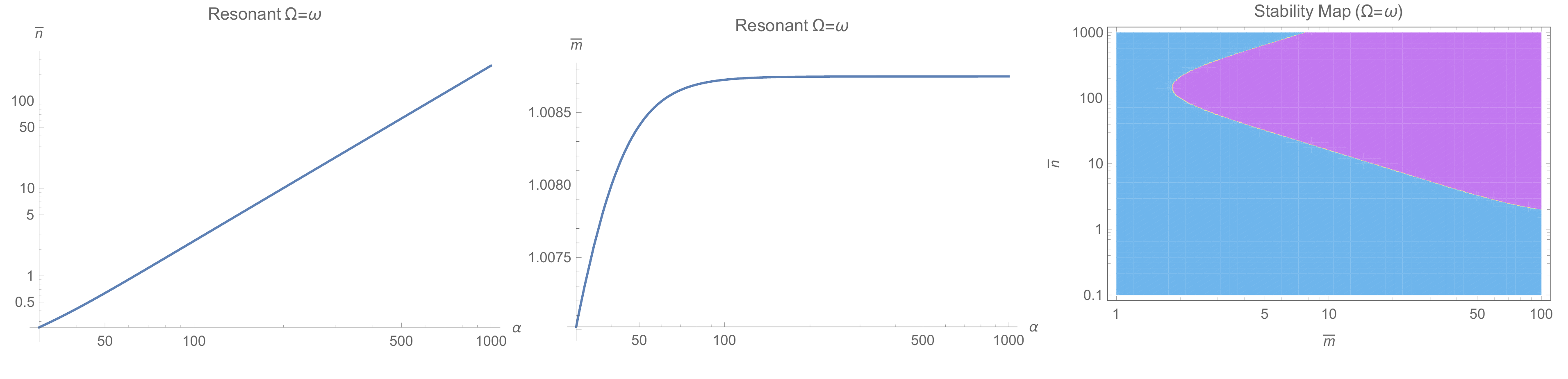}
	\caption{The steady-state photon $\bar{n}$ and phonon $\bar{m}$ population in logarithmic scale, versus input photon flux $\alpha$ at the resonant case $\Omega=\omega$. The right figure shows the stability diagram in terms of $(\bar{n},\bar{m})$ where cyan color represents dynamically stable domains. For the range $(\bar{n},\bar{m})$ on the left, system is unconditionally stable. \label{Fig1ij}}
\end{figure}

The case of resonant excitation with $\omega=\Omega$ exhibits a very different behavior, as illustrated in Fig. \ref{Fig1ij}. One could notice that this time, only the photon population $\bar{n}$ increases almost linearly with the input power with no theoretically ideal limit, while phonon population $\bar{m}$ gets saturated around $\bar{m}\approx 1$ above roughly ten times the same threshold of $10\alpha_\text{sat}$. Quite obviously, too much photon population will cause heating and shift in the working temperature, which respectively will cause a temperature-induced drift in all working parameters. This effect is partially studied in a previous article \cite{Paper3} for zero-point induced spring effect, which is beyond the purpose of current study. However, dynamical stability maps, as shown in Figs. \ref{Fig1cd}-\ref{Fig1ij} reveal no serious concern in the regimes of interest and all systems are unconditionally stable under the studied range of parameters.


\begin{figure}[ht!]
	\centering
	\includegraphics[width=4.8in]{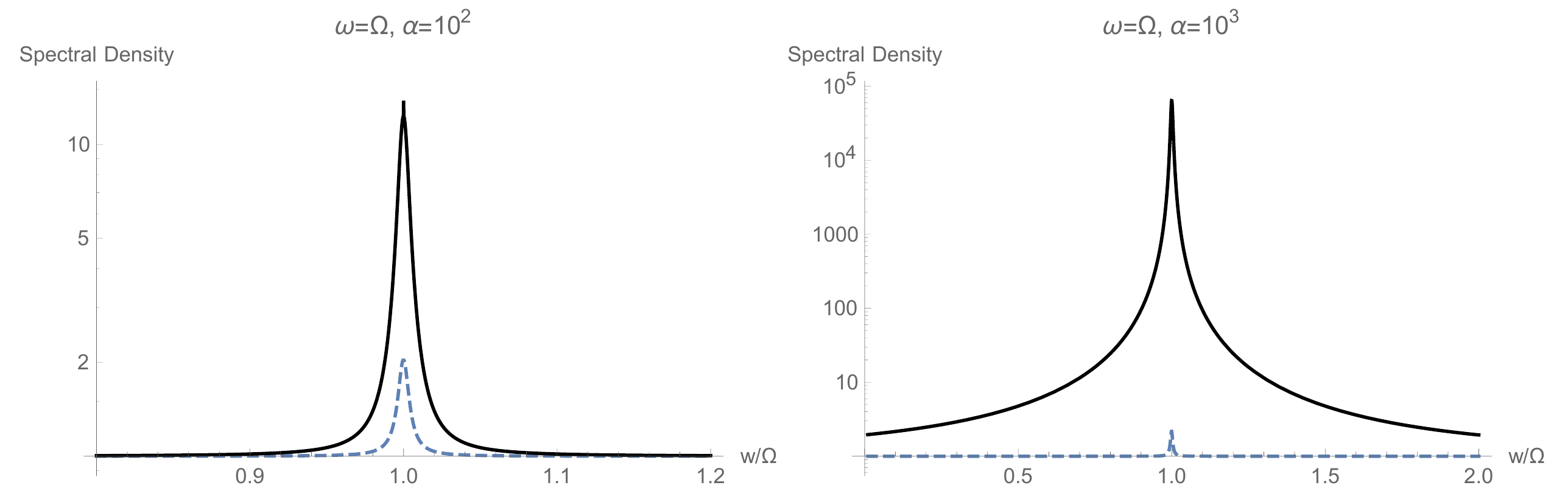}
	\caption{Calculated spectral density $S_{CC}(w/2)$ with (solid black) and without (dashed blue) momentum interaction $\mathbb{H}_2$, for the resonant case with $\omega=\Omega$.\label{Fig2}}
\end{figure}

\begin{figure}[ht!]
	\centering
	\includegraphics[width=4.8in]{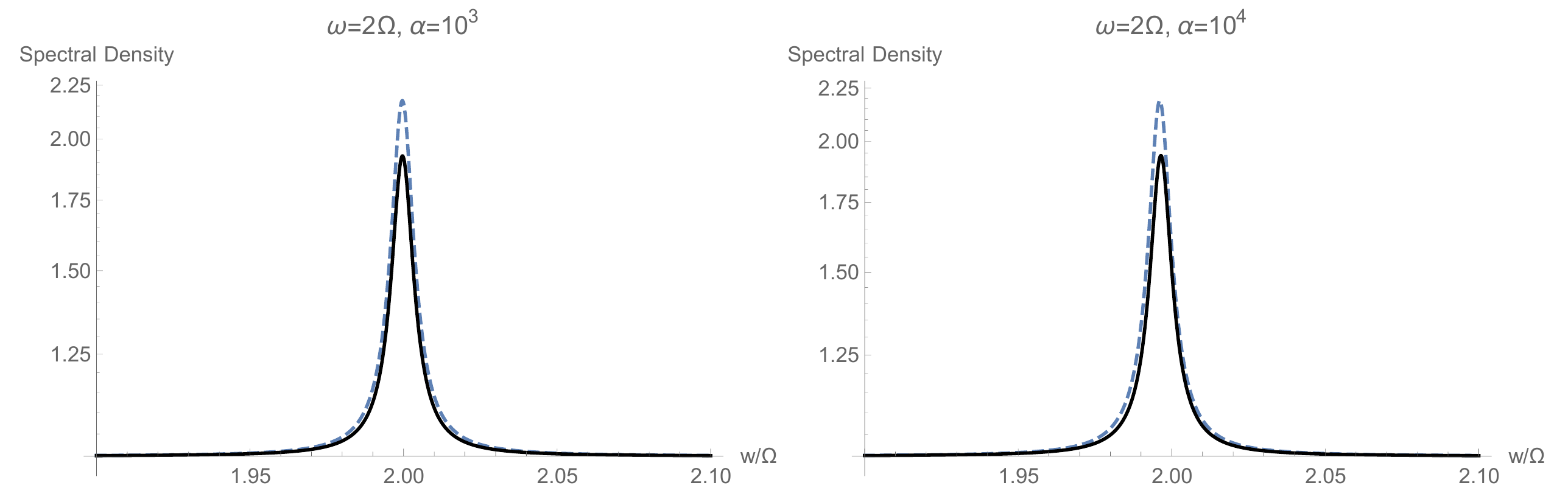}
	\caption{Calculated spectral density $S_{CC}(w/2)$ with (solid black) and without (dashed blue) momentum interaction $\mathbb{H}_2$, for the non-resonant case with $\omega=2\Omega$. \label{Fig3}}
\end{figure}

\begin{figure}[ht!]
	\centering
	\includegraphics[width=4.8in]{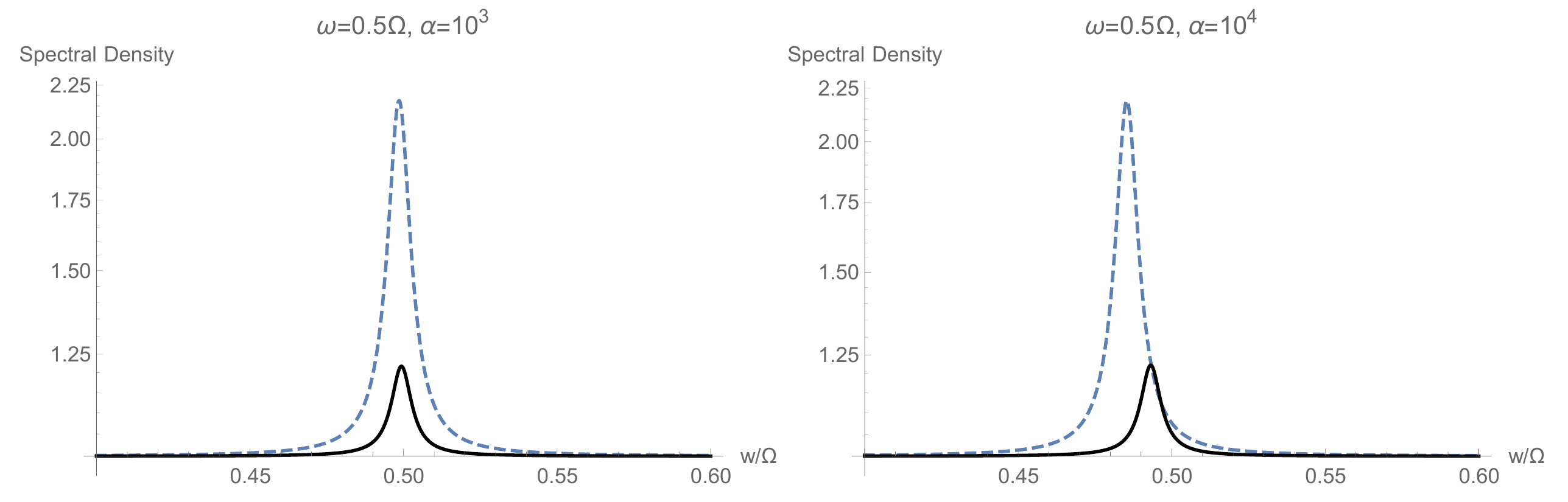}
	\caption{Calculated spectral density $S_{CC}(w/2)$ with (solid black) and without (dashed blue) momentum interaction $\mathbb{H}_2$, for the non-resonant case with $2\omega=\Omega$. \label{Fig4}}
\end{figure}

The next sets of plots in Figs. \ref{Fig2}-\ref{Fig4} show the calculated spectral density $S_{CC}(w/2)$, respectively for resonant and non-resonant cases $\omega=\Omega$, $\omega=2\Omega$, and $2\omega=\Omega$. The frequency axis is already compressed by a factor of 2 because of (\ref{eq21}), which estimates such a halving. In order to study the effect of momentum interaction term $\mathbb{H}_2$, which its existence is predicted in an earlier study \cite{Paper1}, we estimate the spectral densities under two conditions. First, we exclude the momentum interaction and only keep the standard quadratic term $\mathbb{H}_1$. Curves plotted in dahsed blue demonstrated this regime of having only standard quadratic interaction. Curves plotted in solid black correspond to simulation results where the non-standard quadratic interaction with momentum conservation $\mathbb{H}_2$ has been included. We observe here the markedly prominent effect of $\mathbb{H}_2$ under resonant case with high drive $\alpha=10^3\text{Hz}$ shown in Fig. \ref{Fig2}. At lower intensites, spectral responses differ in magnitude and $\mathbb{H}_2$ has caused an intensifying in quadratic interactions. This significant change in optical spectral response across the mechanical frequency at $\omega=\Omega$ can be taken as a primary signature of existence of momentum interaction term $\mathbb{H}_2$, predicted earlier \cite{Paper1}. 

The behavior at optical frequencies $\omega>\Omega$ is quite expected. Since from (\ref{eq2}) we have $\beta/\varepsilon\propto(\Omega/\omega)^2$, one would estimate that at increased optical frequencies, the effect of non-standard interaction $\mathbb{H}_2$ would start to fade out. This is quite correct, indeed, as at high optical frequencies $\omega>>\Omega$, we notice the gradually disappearing effect of non-standard quadratic term $\mathbb{H}_2$, as illustrated in Fig. \ref{Fig3} for $\omega=2\Omega$. When larger values for the ratio $\omega/\Omega$ is selected, the effect of $\mathbb{H}_2$ is completely wiped out. This fact also confirms why the non-standard momentum interaction $\mathbb{H}_2$ has not been observed in experiments so far, as all of the quadratically studied optomechanical interactions fall in this regime of $\omega>>\Omega$. Also, we may identify the approximate limit of $(\omega/\Omega)^2\leq 10$ beyond which there would be no observable non-standard quadratic interaction.

However, when the mechanical frequency gets large with respect to the optical frequency $\Omega>\omega$, the effect of non-standard interaction $\mathbb{H}_2$ becomes quite evident. This has been shown in Fig. \ref{Fig4}. First of all, momentum interactions tend to reduce the overall quadratic interaction under studied conditions. Secondly, at higher illumination and photon flux rates, we notice a clear displacement of the peak toward higher frequencies because of the non-standard quadratic interaction $\mathbb{H}_2$. This fact can serve as a secondary and unmistakable signature of existence of momentum interaction terms in an experiment.

\begin{figure}[ht!]
	\centering
	\includegraphics[width=4.8in]{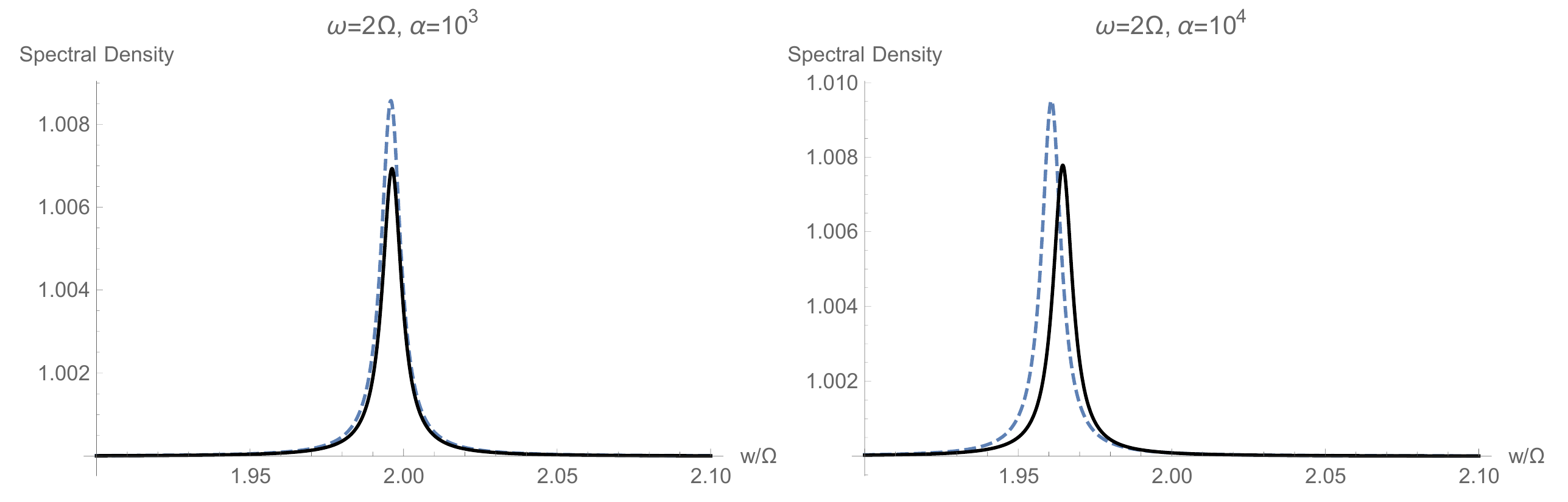}
	\caption{Calculated spectral density $S_{CC}(w/2)$ with (solid black) and without (dashed blue) momentum interaction $\mathbb{H}_2$, for the non-resonant case with $\omega=2\Omega$ and strong coupling $\varepsilon=2\pi\times 500{\rm Hz}$. \label{Fig5}}
\end{figure}

\begin{figure}[ht!]
	\centering
	\includegraphics[width=4.8in]{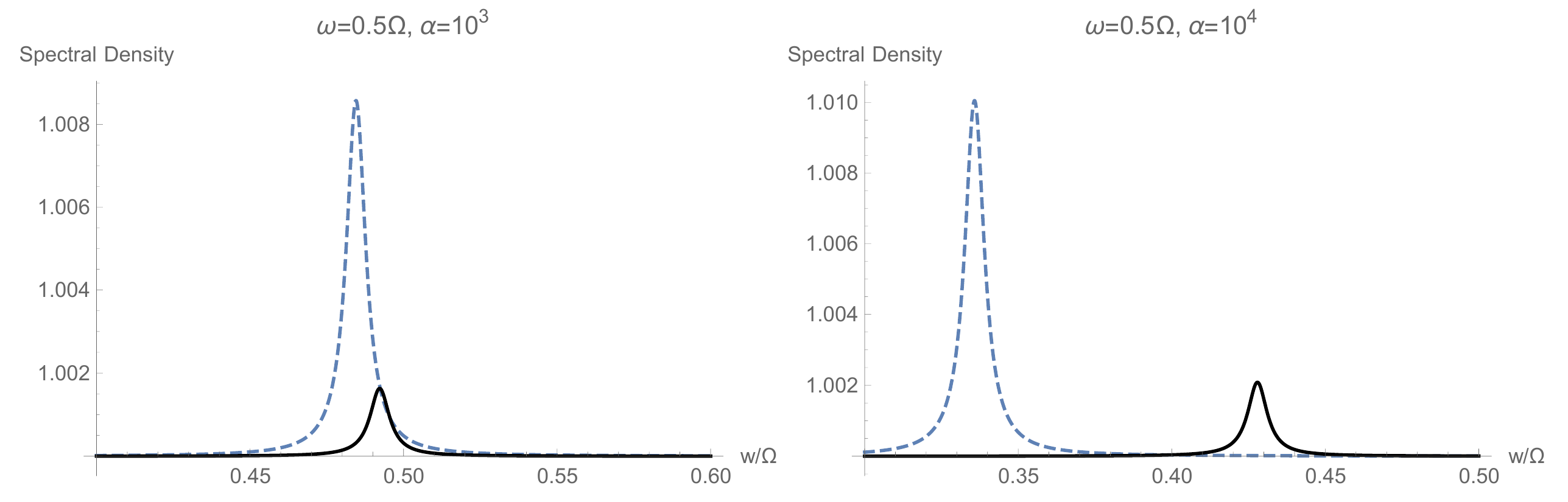}
	\caption{Calculated spectral density $S_{CC}(w/2)$ with (solid black) and without (dashed blue) momentum interaction $\mathbb{H}_2$, for the non-resonant case with $2\omega=\Omega$ and strong coupling $\varepsilon=2\pi\times 500{\rm Hz}$. \label{Fig6}}
\end{figure}

\begin{figure}[ht!]
	\centering
	\includegraphics[width=4.8in]{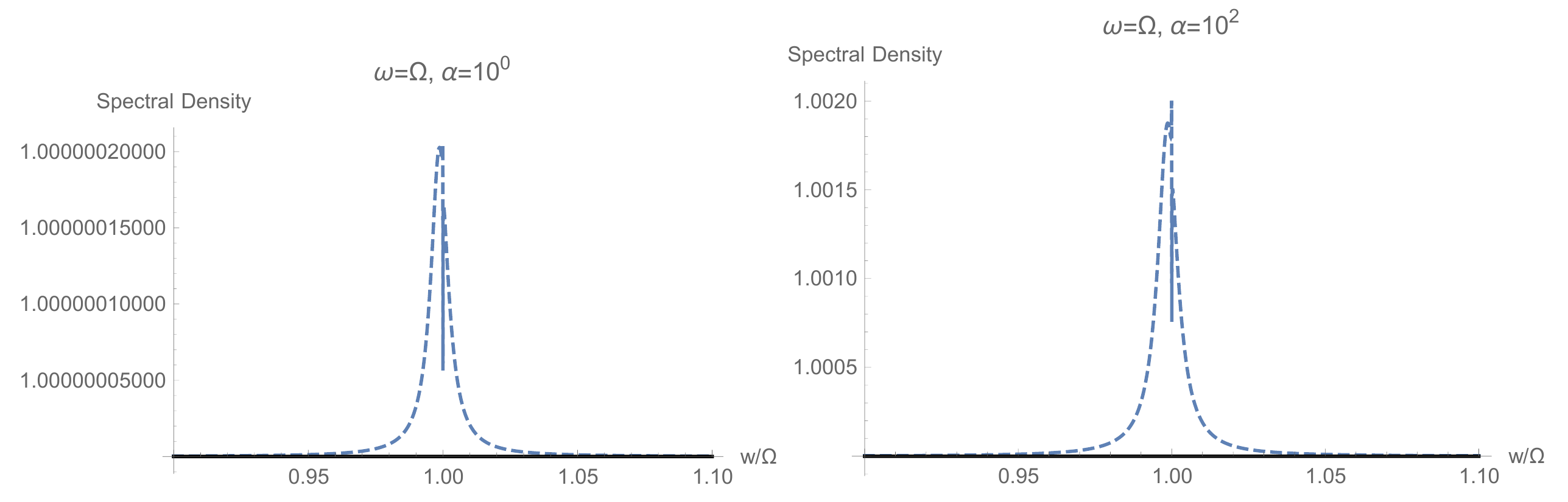}
	\caption{Calculated spectral density $S_{CC}(w/2)$ with (solid black) and without (dashed blue) momentum interaction $\mathbb{H}_2$, for the resonant case with $\omega=\Omega$ and strong coupling $\varepsilon=2\pi\times 500{\rm Hz}$. \label{Fig7}}
\end{figure}

Next, we turn to the case of strongly coupled quadratic optomechanics with $\varepsilon=2\pi\times 500{\rm Hz}$. This regime has been studied in the next sets of plots included in Figs. \ref{Fig5}-\ref{Fig7}. This time, the effect of strong coupling on the quadratic response is much pronounced, as opposed to the weak coupling. We notice that the behavior of steady state photon $\bar{n}$ and phonon $\bar{m}$ populations is essentially similar to those of weakly coupled regime, with the difference that the flux threshold for saturation or increase is shifted to the higher value of roughly $\alpha_\text{sat}=200\text{Hz}$ by an order of magnitude. The corresponding graphs are not shown for the sake of brevity. Again, one could observe that at higher optical frequencies $\omega>\Omega$, the effect of momentum exchange term $\mathbb{H}_2$ is much smaller than the standard quadratic interaction $\mathbb{H}_1$. This fact is shown in plots of Fig. \ref{Fig5}. Meanwhile, at small optical frequencies where mechanical frequency prevails $\Omega>\omega$, non-standard interactions $\mathbb{H}_2$ can have a dominant effect on the response. This is more or less similar to the weak coupling regime show in Fig. \ref{Fig3}. However, the overall spectral response is actually less intense, since the associated peaks are smaller. Hence, increasing the strength of quadratic interaction does not necessarily increase the intensity of exhibited response. Furthermore, one could notice a marked shift in spectrum toward higher frequencies at sufficiently high photon input flux.

The case of large mechanical frequency $\Omega>\omega$ under the strong coupling regime is also studied in Fig. \ref{Fig6}, as opposed to the weak coupling regime in Fig. \ref{Fig4}. The general trend of behavior from weak to strong coupling regimes at $\alpha=10^3\text{Hz}$ has not essentially changed, and there is an expected shift in the peak in spectrum toward higher frequencies because of non-standard interactions. However, continuing to increase the input optical intensity causes complete and distinct separation among the corresponding peaks, while both move toward lower frequencies. This reveals the fact that a tertiary signature of momentum exchange term $\mathbb{H}_2$ can be taken as the central location of peak in response, which is dependent on the possible existence of $\mathbb{H}_2$.

\begin{figure}[ht!]
	\centering
	\includegraphics[width=4.8in]{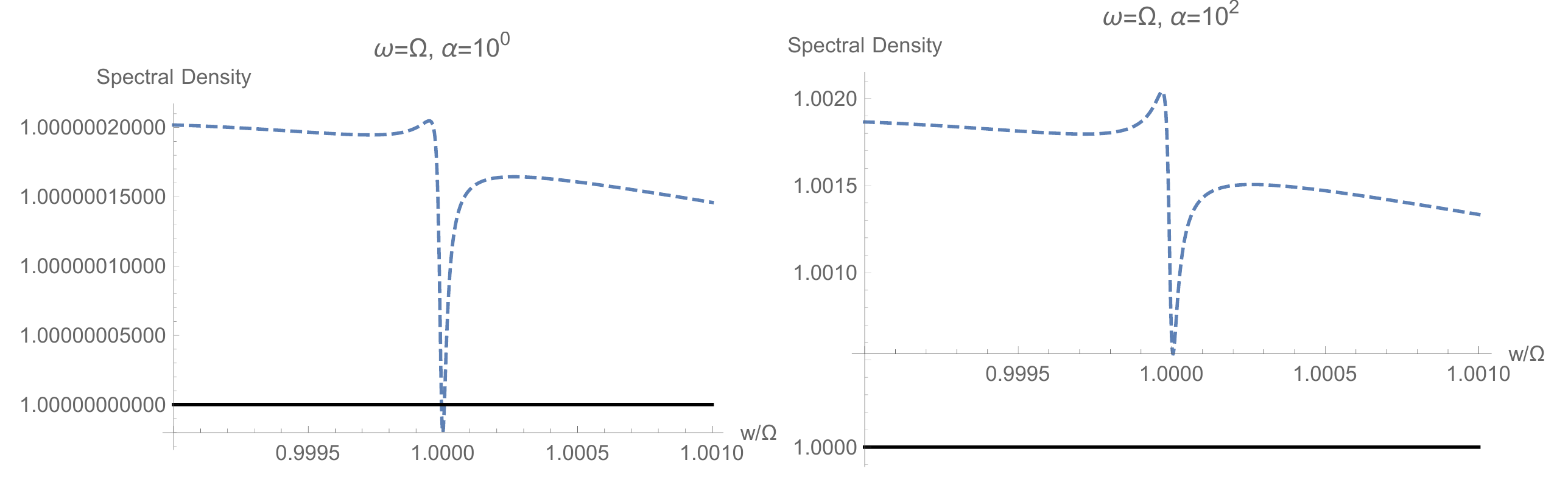}
	\caption{Behavior of sharp resonances at $w=\omega$ with (solid black) and without (dashed blue) momentum interaction $\mathbb{H}_2$, for the resonant case with $\omega=\Omega$ and strong coupling $\varepsilon=2\pi\times 500{\rm Hz}$. The non-standard quadratic interaction completely wipes out the sharp resonance and peak in the spectrum. \label{Fig8}}
\end{figure}

Finally, we studied the resonant case $\omega=\Omega$ under strong coupling, shown in Fig. \ref{Fig7} as opposed to the results of weak coupling shown in Fig. \ref{Fig2}. We notice a very distinct difference between these two cases. It can be observed that under strong coupling, the standard quadratic response is expected to exhibit a very sharp resonance exactly at $w=\omega$, which is actually placed a bit higher than the central peak frequency of the output spectrum. The existence of these resonances in clearly shown by zoomed out plots of Fig. \ref{Fig8}. One could tell, that existence of a tiny squeezing even could have been expected under standard quadratic interaction $\mathbb{H}_1$, which is again entirely wiped out by the overriding presence of non-standard momentum interaction term $\mathbb{H}_2$. This can be taken as the existence of quaternary signature of momentum-field interactions, which eliminates such squeezing and sharp resonances from the spectral response. The behavior under ultrastrong coupling has been calculated studied, but since it is essentially similar to the strongly coupled case, the results are not shown and dropped from the article. 

The results of this research could be relevant to sensing applications where nonlinear optomechanics could in principle produce pronounced sensitivities, such as those reported recently for quantum gravimetry \cite{q14,Thesis}.  Also possible generation of continuous squeezed electromagnetic radiation seems to be feasible by carefully optimized design of a membrane-in-the-middle superconducting electromechanics setup. Further applications could be still plausible but will need more in depth study. Experiments can unambiguously determine the existence of non-standard quadratic optomechanics.

\section*{Conclusions}

In summary, we presented a detailed theoretical and numerical analysis of the quadratic optomechanical interactions, using the method of higher-order operators. We have studied both types of standard quadratic and the predicted non-standard quadratic interactions, and through extensive simulations have established clear signatures for existence of non-standard interactions, while stability of the system under study has also been established. Such types of quadratic interactions can be probed in a carefully designed experiment where the optical and mechanical frequencies fall within the same order of magnitude. 

\section*{Acknowledgments}
	
Stimulating discussions of this work with Dr. David Edward Bruschi at the University of Vienna, Prof. Radim Filip at Palack\'{y} University, members of the Laboratory of Quantum Foundations and Quantum Information at Nano- and Micro-scale in Vienna Center for Quantum Science and Technology (VCQ), as well as members of the Laboratory of Photonic and Quantum Measurements at \'{E}cole Polytechnique F\'{e}d\'{e}rale de Lausanne (EPFL) is much appreciated. This work is dedicated to the celebrated artist, Anastasia Huppmann.
	
\section*{Additional Information}
The author declares no competing interests.

\end{document}